\documentclass[12pt,aps,prd]{revtex4}
\tighten
\usepackage{latexsym}
\usepackage{epsfig}

\newcommand{\g}{\gamma}

\newcommand{\be}{\begin{equation}}
\newcommand{\ee}{\end{equation}}
\newcommand{\gsim}{\raise.3ex\hbox{$>$\kern-.75em\lower1ex\hbox{$\sim$}}}
\newcommand{\lsim}{\raise.3ex\hbox{$<$\kern-.75em\lower1ex\hbox{$\sim$}}}
\newcommand{\ba}{\begin{eqnarray}}
\newcommand{\ea}{\end{eqnarray}}
\newcommand{\bea}{\begin{eqnarray}}
\newcommand{\eea}{\end{eqnarray}}
\newcommand{\bean}{\begin{eqnarray*}}
\newcommand{\eean}{\end{eqnarray*}}

\newcommand{\bk}{{\bf k}}

\def\spose#1{\hbox to 0pt{#1\hss}}
\def\ltapprox{\mathrel{\spose{\lower 3pt\hbox{$\mathchar"218$}}
\raise 2.0pt\hbox{$\mathchar"13C$}}}
\def\gtapprox{\mathrel{\spose{\lower 3pt\hbox{$\mathchar"218$}}
\raise 2.0pt\hbox{$\mathchar"13E$}}}

\newcommand{\SM}{{\cal Q}}
\newcommand{\QQ}{Q}

\begin{document}

%\draft
%\preprint{\
%\begin{tabular}{rr}
%CfPA/96-th-15 &  \\
%&
%\end{tabular}
%}

%\twocolumn[\hsize\textwidth\columnwidth\hsize\csname@twocolumnfalse\endcsname

\title{Cosmological perturbations from varying masses and couplings}

\author{Filippo Vernizzi}
\email{vernizzi@iap.fr} \affiliation{Institut d'Astrophysique de
Paris, GR$\varepsilon$CO, FRE 2435-CNRS, 98bis boulevard Arago,
75014 Paris, France}

%\maketitle

\vspace{1cm}

\begin{abstract}
We study the evolution of perturbations during the domination and
decay of a massive particle species whose mass and decay rate are
allowed to depend on the expectation value of a light scalar
field. We specialize in the case where the light field is
slow-rolling, showing that during a phase of inhomogeneous
mass-domination and decay the isocurvature perturbation of the
light field is converted into a curvature perturbation with an
efficiency which is nine times larger than when the mass is fixed.
We derive a condition on the annihilation cross section and 
on the decay rate for the
domination of the massive particles and we show that standard
model particles cannot dominate the universe before
nucleosynthesis. We also compare this mechanism with the curvaton
model. Finally, observational signatures are discussed. A cold dark matter
isocurvature mode can be generated if the dark matter is produced
out of equilibrium by both the inflaton and the massive particle
species decay. Non-Gaussianities are present: they are chi-square
deviations. However, they might be too small to be
observable.

\end{abstract}

\maketitle

\date{\today}
%\pacs{PACS Numbers : 98.80.Cq, 98.70.Vc, 98.80.Hw}

\renewcommand{\thefootnote}{\arabic{footnote}} \setcounter{footnote}{0}

\section{Introduction}

Strongly supported by the recent observations of 
the Wilkinson Microwave Anisotropy Probe (WMAP) 
satellite \cite{WMAP1,WMAP2,Peiris,WMAP4,WMAP5}, inflation
\cite{guth81} has now become the dominant contender for generating
adiabatic density perturbations with an almost flat spectrum. In
the standard picture, the observed large scale density
perturbations are due to the fluctuations of the inflaton field,
which are created during a period of accelerated expansion by
amplification of quantum vacuum fluctuations \cite{MFB}.

Recently, alternative mechanisms for generating density
perturbations after inflation have been proposed. They all assume
that the early universe is filled with at least one light scalar
field $\phi$ whose energy density is negligible during inflation.
Fluctuations in the field are amplified during the inflationary
phase with a quasi scale-invariant spectrum and their amplitude
fixed by the energy scale of inflation, $\delta \phi \sim
\frac{H_*}{2\pi}$, where $H_*$ is the Hubble parameter at the horizon
crossing. If the vacuum expectation value (VEV) of the light field
$\phi$ during inflation is small, its density perturbation $\delta
\rho_\phi /\rho_\phi \sim \delta \phi/\phi$ is larger than the
perturbations generated during inflation. However, since the light
field is sub-dominant, its perturbation is initially of the
isocurvature type. Later, it can be converted into a curvature
perturbation. For a successful conversion one of these two
ingredients are necessary:  (1) the field $\phi$ must come (close)
to dominate the universe, or (2) it must induce fluctuations in a
second component which eventually comes to dominate the universe.

In case (1) the scalar field $\phi$ is called curvaton.
In the curvaton scenario
\cite{curvaton1,curvaton2,curvaton3}, the isocurvature
perturbation stored in the curvaton field is transformed into
a curvature perturbation during a phase in which $\phi$ oscillates
at the bottom of its potential behaving as a dust-like
component and thus dominating over the radiation.

Case (2) has been inspired by the idea that coupling constants and
masses of particles during  the early universe may depend on the
VEV of some light scalar field. This idea is motivated by
supersymmetric theories and theories inspired by superstrings
where coupling `constants' and masses of particles are usually
functions of the scalar fields of the theory. An interesting
proposal is that large scale perturbations could be generated from
the fluctuations of the inflaton coupling to ordinary matter
\cite{gamma1,gamma2}. These can be converted into curvature
perturbations during the reheating phase \cite{Turner} when the
inflaton decays into radiation and reheats different patches of
the universe with different temperatures and energy densities (see
however \cite{Enqvist}). This mechanism of conversion has been
called inhomogeneous reheating in \cite{gamma1}. An extension and
generalization of this mechanism is that the masses of particles
produced during the reheating are allowed to vary \cite{gamma3}.
If sufficiently long-lived, the massive particles can eventually
dominate the universe. Due to the fluctuations in the masses the
mass-domination process becomes inhomogeneous and can convert the
density perturbation of $\phi$ into a curvature perturbation.

Some authors \cite{Riotto,Mazum,Tsuji} (see also \cite{Zalda})
have studied the evolution of large scale perturbations during the
decay of the inflaton in the case of a fluctuating decay rate. Two
of these groups \cite{Riotto,Mazum} make extensive use of the
formalism introduced in Ref.~\cite{wmu} (see however also
\cite{KS,hama}) for interacting fluids. Indeed, on several
occasions the light scalar field can be treated as a fluid and the
formalism of Ref.~\cite{wmu} consistently applied. In this paper
we extend the analysis of \cite{Riotto,Mazum} and we include the
case where the inhomogeneous reheating is due to the decay of
massive particles whose mass and decay rate depend on the VEV of a
light field. In the limit of a constant mass the results of
\cite{Riotto,Mazum} are recovered.

We also derive the condition for the domination of the massive
particles. Since at early times they are in thermal equilibrium,
their abundance depends on their annihilation cross section. As we shall
see the condition for the domination does not depend on their mass
but only on their annihilation cross section and decay rate. Our results are
then compared with those obtained for the curvaton model and some
of the observational consequences are discussed. If the inflaton
and/or the massive particles decay out of equilibrium isocurvature
perturbations can be generated. The inhomogeneous inflaton decay
and mass-domination can also lead to non-Gaussianities in the
adiabatic spectrum of perturbations. These can lead to
observational signatures in the cosmic microwave background (CMB)
that can be possibly observed in future experiments.

The paper is organized as follows. In the next section we model the
inhomogeneous mass-domination mechanism
by writing down the conservation laws for the radiation, the massive
particles and the light field, and we derive the coupling
between these species.
In Section III we write the background and perturbation
equations obtained from the conservation laws. These equations
are completely general. They can be used for the inhomogeneous inflaton decay
as for the inhomogeneous mass-domination model.
In Section
IV we concentrate on the limit that the light
field is in slow-roll and we discuss the efficiency of the inhomogeneous
inflaton decay and mass-domination models. We also compare them to the
curvaton model. Finally, in Section V we discuss the observable signatures
of the inhomogeneous mass-domination i.e., the possible presence of
isocurvature perturbations and
non-Gaussianities.
In the final section we draw our conclusions.

\section{Modeling the couplings}

As mentioned in the Introduction, we want to derive a set of
equations that, depending on the initial conditions, can describe
two different physical situations: (A) inhomogeneous inflaton
decay or (B) inhomogeneous mass-domination and decay. In order to
describe these two cases we consider a universe containing three
components: a radiation fluid $\g$, a non-relativistic fluid
$\psi$, and a light scalar field $\phi$. According to the
different physical situations we have:
\begin{itemize}
\item[(A)] Inhomogeneous inflaton decay: At the end of inflation
the  universe is dominated by the inflaton. This coherently
oscillates at the bottom of its potential behaving as a
non-relativistic fluid $\psi$. The decay rate of the inflaton,
$\Gamma$, depends on $\phi$. \item[(B)] Inhomogeneous
mass-domination: After inflation and reheating, the universe is
filled with radiation $\g$ and with a species $\psi$ of
non-relativistic massive particles, whose mass $M$ and decay rate
$\Gamma$ depend on $\phi$.
\end{itemize}
As we shall see, the equations derived for case (B) can be applied
to case (A) in the limit that $M$ does not depend on $\phi$.
Therefore, in the following we  concentrate on the inhomogeneous
mass-domination and decay.

The universe is initially
dominated by a radiation bath $\gamma$ with
temperature $T$ and by a
massive particle species with mass $M \gtapprox T$, such that
it can be described as a non-relativistic fluid $\psi$.
If the massive particles froze out at
a temperature such that $M/T$ was not much larger
than 1, then the species can have a significant
relic abundance and, if sufficiently long-lived,
can eventually dominate the universe before
decaying. The condition for the domination can be
obtained by requiring the decay rate to be smaller than the
Hubble parameter at the moment when $\psi$ decays. The latter corresponds
approximately to $\sim M^2/M_{\rm Pl}$ if we assume that $\psi$ starts
to
dominate as soon as it becomes non-relativistic, $T \sim M$.
Thus we find  \cite{gamma3},
\be
\Gamma < M^2/M_{\rm Pl}, \label{eq:conddomin}
\ee
where $M_{\rm Pl}\equiv (8 \pi G)^{-1/2}$
is the reduced Planck mass. A more precise condition for the
domination is derived
in Sec.~IV.B and depends only on the annihilation cross section 
and decay rate of the
massive particles.
When the $\psi$-particles decay into $\g$
they release a considerable amount of
entropy and reheat the universe once again.
During the inhomogeneous mass-domination and reheating,
fluctuations in $M=M(\phi)$ and $\Gamma=\Gamma(\phi)$
can be converted into
curvature perturbations.

Variations of masses and coupling constants can be easily
motivated in the context of string theory where compactified extra
dimensions emerge as massless scalar fields, called moduli, in the
effective four dimensional theory. These moduli  generally couple
directly to matter leading to variations of masses and fundamental
constants \cite{Damour}. That the mass of particles may depend on
a light field has been discussed in several places in the
literature, especially in the context of interacting dark matter
and dark energy (see  e.g., \cite{Peebles} for a review) and more
recently in models with variation of the fine structure constant
\cite{Mota} and in the chameleon cosmology \cite{Khoury}. A field
dependent mass leads to a non-trivial coupling between the field
and the non-relativistic fluid made of massive particles. The form
of this coupling can be derived by considering the
non-relativistic fluid as a classical gas of point-like particles
with mass $M(\phi)$ and action $S = - \int M ds$ \cite{Damour}. In
this section we derive the conservation equations of our three
fluids system using the conservation of the total energy.

Let $T_{\mu \nu}^{(\gamma)}$, $T_{\mu \nu}^{(\psi)}$, and
$T_{\mu \nu}^{(\phi)}$ be the energy-momentum
tensors of the three components $\g$, $\psi$, and $\phi$.
In particular we write the energy-momentum tensor of the
non-relativistic massive species as
\be
T_{\mu \nu}^{(\psi)} =\rho_\psi u_\mu u_\nu = M n_\psi u_\mu u_\nu,
\ee
where $\rho_\psi$ is the energy density and $n_\psi$ the number density of
$\psi$.
From general covariance we require that the sum of the three
energy-momentum tensors is conserved,
\bea
0&=& \nabla_\mu (T_{ \nu}^{(\phi) \mu} + T_{ \nu}^{(\g) \mu}
+ T_{ \nu}^{(\psi) \mu}) \\
 &=&
\nabla_\mu T_{ \nu}^{(\phi) \mu} + \nabla_\mu T_{ \nu}^{(\g) \mu}
+ \frac{\partial_\mu M}{M}T_{ \nu}^{(\psi) \mu} + M  \nabla_\mu
(n_\psi u^\mu) u_\nu + M n_\psi u^\mu \nabla_\mu u_\nu .
\label{eq:summa} \eea The last term of the right hand side of this
equation vanishes only if the massive particles follow a geodesic,
which is not the case if they interact with other fluid
components. We project Eq.~(\ref{eq:summa}) on the unit vector
$u^\nu$ and we define the decay rate of $\psi$ as the rate of
change of the number of $\psi$-particles, \be \nabla_\mu (n_\psi
u^\mu) = -\Gamma n_\psi. \ee By assuming that $\psi$ decays only
into $\g$, from Eq.~(\ref{eq:summa}) we obtain three coupled
conservation equations, \bea u^\nu \nabla_\mu T_{ \nu}^{(\psi)
\mu} &=&  - ( C_M \partial_\mu \phi u^\mu - \Gamma)
\rho_\psi , \label{eq:1}\\
u^\nu \nabla_\mu T_{ \nu}^{(\gamma) \mu} &=&  -\Gamma \rho_\psi , \\
u^\nu \nabla_\mu T_{ \nu}^{(\phi) \mu} &=& C_M \partial_\mu \phi
u^\mu \rho_\psi, \label{eq:2} \eea where we have defined the mass
coupling function $C_M$, \be C_M(\phi) \equiv  \frac{\partial \ln
M}{\partial \phi}. \ee For later convenience, we also define the
decay rate coupling function $C_\Gamma$, \be C_\Gamma(\phi) \equiv
\frac{\partial \ln \Gamma}{\partial \phi}. \ee
Equations~(\ref{eq:1}-\ref{eq:2}) will be used to derive the background
and perturbation evolution equations describing the inhomogeneous
mass-domination. The perturbation equations for the coupled
field-fluid system ($\Gamma=0$) in the case of a linear coupling
($C_M=$ constant) have been also discussed in \cite{Amendola} in
the context of coupled quintessence. Finally, if $C_M=0$,
Eqs.~(\ref{eq:1}-\ref{eq:2}) describe the inhomogeneous inflaton
decay, as discussed in \cite{Riotto,Mazum}.

\section{Cosmological perturbations}

Here we write down and discuss the background and perturbation
equations derived from Eqs.~(\ref{eq:1}-\ref{eq:2}).

\subsection{Background equations}

Consider a Friedmann-Lema\^{\i}tre  universe, with metric
$ds^2 = -dt^2+a^2(t) dx^2$,  governed by the Friedmann
equations,
\bea
\label{eq:Friedmann}
H^2 &=& \frac{1}{3M^2_{\rm Pl}} \rho
\equiv \frac{1}{3M^2_{\rm Pl}} (\rho_\psi + \rho_\gamma + \rho_\phi ), \\
\dot H &=& -\frac{1}{2M^2_{\rm Pl}} (\rho +P) \equiv
-\frac{1}{2M^2_{\rm Pl}}
\left( \rho_\psi + \frac{4}{3} \rho_\gamma  + \dot{\phi}^2 \right),
\eea
where $\rho$ and $P$ are the total energy density and pressure
of the universe, $\rho_\gamma$ and $\rho_\phi$ are the
energy density of the $\gamma$ and $\phi$ species, respectively,
and $H$ is the Hubble parameter, defined as $H \equiv \dot a /a$.

The evolution equation governing the total energy density and
ensuring the conservation of the total energy is
\be
\dot \rho = - H \left(3 \rho_\psi + 4 \rho_\gamma +3 \dot \phi^2 \right).
\ee
In particular we have three coupled conservation equations
which can be derived from
Eqs.~(\ref{eq:1}-\ref{eq:2}). They read
\bea
\label{eq:dyn1}
\dot{\rho}_\psi &=& - (3H - C_M \dot \phi + \Gamma)  \rho_\psi , \\
\label{eq:dyn2}
\dot{\rho}_\gamma &=& - 4 H \rho_\gamma + \Gamma \rho_\psi, \\
\label{eq:dyn3}
\dot{\rho}_\phi &=& -3 H  \dot \phi^2
- C_M \dot \phi \rho_\psi.
\label{eq:dyn4}
\eea
In order to solve for $\dot \phi$ we need the
field evolution equation,
\be
\ddot \phi + 3H\dot \phi + V_{, \phi} = - C_M \rho_\psi,
\ee
where $V_{, \phi}$ is the derivative of
the potential of $\phi$, $V=V(\phi)$, with respect to the field $\phi$.

As first introduced in \cite{KS} we define the energy transfer
functions $\QQ_{\alpha}$ (with $\alpha=\psi,\gamma,\phi$) as \be
\QQ_\alpha \equiv \dot \rho_\alpha + 3 H (\rho_\alpha + P_\alpha)
. \ee From Eqs.~(\ref{eq:dyn1}-\ref{eq:dyn4}) we have
\begin{eqnarray}
\label{eq:defbackQa}
\QQ_{\psi} &=& ( C_M \dot \phi -  \Gamma) \rho_{\psi} ,  \\
\label{eq:defbackQb}
\QQ_{\gamma} &=& \Gamma \rho_{\psi} , \\
\QQ_{\phi} &=& - C_M \dot \phi \rho_\psi.\label{eq:defbackQc}
\end{eqnarray}

\subsection{Perturbation equations}

Here we discuss the perturbation equations derived from
Eqs.~(\ref{eq:1}-\ref{eq:2}).
We describe scalar perturbations in the metric with
the line element
\begin{equation}
ds^2=-(1+2\Phi)dt^2
+a^2 (1-2\Psi)\delta_{ij} dx^idx^j ,
\end{equation}
where $\Phi$ and $\Psi$ correspond to the Bardeen potentials in
longitudinal gauge. In the absence of anisotropic stress perturbation
$\Phi = \Psi$.
In order to perturb the energy-momentum tensor of the three components
$\psi$, $\g$, and $\phi$, we introduce the energy density and pressure
perturbations,  $\delta \rho_\alpha$ and
$\delta P_{\alpha}$.

For  each fluid component $\alpha$
one can introduce the gauge invariant
curvature perturbation on the uniform
$\alpha$-energy density hypersurface defined as
\begin{eqnarray}
\zeta_{\alpha} &\equiv& -\Psi-H
\frac{\delta\rho_\alpha}{\dot{\rho_\alpha}}.
\end{eqnarray}
The $\zeta_\alpha$'s remain constant on large scales only for
adiabatic perturbations and in any fluid whose energy-momentum
tensor is locally conserved: $n^\nu T_\nu^{(\alpha) \mu}=0$
\cite{WMLL}. In our case $\QQ_{\alpha} \neq 0$ and this is not the
case. The total uniform curvature perturbation, introduced by
Bardeen \cite{Bardeenfirst} and Bardeen, Steinhardt and Turner
\cite{BST}, is given by a weighted sum of the individual uniform
curvature perturbations, \be
 \zeta \equiv -\Psi - H\frac{\delta
\rho}{\dot\rho}= \sum_\alpha \frac{\dot \rho_\alpha}{\dot \rho}
\zeta_\alpha,  \label{eq:zetatot}
\end{equation}
and the relative entropy perturbation between two components
$\alpha$ and $\beta$ is given by the difference
between the two uniform curvature perturbations $\zeta_\alpha$
and $\zeta_\beta$ \cite{wmu},
\be
 \label{eq:defS}
  {\cal S}_{\alpha \beta} \equiv 3(\zeta_\alpha-\zeta_\beta).
\end{equation}

According to Ref.~\cite{Garcia}, at large scales (neglecting
spatial gradients) the time evolution of the curvature
perturbation can be written as \be \dot \zeta = - \frac{H}{\rho
+P} \delta P_{\rm nad}, \label{eq:totalzetaevol} \ee where $\delta
P_{\rm nad}$ is the total non-adiabatic pressure perturbation,
given by the sum of the intrinsic non-adiabatic pressure
perturbation of each component and the relative non-adiabatic
pressure perturbation, \be \delta P_{\rm nad} = \sum_\alpha \delta
P_{{\rm intr}, \alpha} + \delta P_{\rm rel}, \ee where the
intrinsic non-adiabatic pressure perturbation of the species
$\alpha$ is given as \be \delta P_{{\rm intr}, \alpha} \equiv
\delta P_\alpha - c_\alpha^2 \delta \rho_\alpha, \label{eq:nonad}
\ee and the relative non-adiabatic pressure perturbation depends
on the uniform curvature perturbations, \be \delta P_{\rm rel}
\equiv \sum_\alpha \frac{\dot \rho_\alpha c_\alpha^2}{H} (\zeta -
\zeta_\alpha) = - \frac{1}{6 H \dot \rho} \sum_{\alpha, \beta}
\dot \rho_\alpha \dot \rho_\beta (c_\alpha^2 - c_\beta^2 ) {\cal
S}_{\alpha \beta}. \label{eq:noadrel} \ee Here $c^2_\alpha \equiv
\dot P_\alpha / \dot \rho_\alpha $ is the adiabatic speed of sound
of the species $\alpha$. At large scales, the evolution equation
for each individual uniform curvature perturbation $\zeta_\alpha$
is given by \cite{wmu} \be \dot\zeta_\alpha =
 -{H\left(
 \delta
 \QQ_{\rm{intr},\alpha}+ \delta \QQ_{\rm{rel},\alpha}
-3 H \delta P_{{\rm intr}, \alpha} \right)
\over\dot{\rho_\alpha}} . \label{eq:partialzetaevol}
\ee
The $\zeta_\alpha$'s are sourced by three gauge invariant terms, the
intrinsic and the relative non-adiabatic energy transfer functions,
and the intrinsic non-adiabatic pressure perturbation.

We now describe, one by one, the three terms on the right hand
side of Eq.~(\ref{eq:partialzetaevol}). The non-adiabatic pressure
perturbation $\delta P_{{\rm intr},\alpha}$ is defined in
Eq.~(\ref{eq:nonad}). For a fluid whose parameter of state
$w_\alpha \equiv P_\alpha / \rho_\alpha$ is constant, $c^2_\alpha
= w_\alpha$ and $\delta  P_{{\rm intr}, \alpha}=0$. This is the
case for the non-relativistic species $\psi$ and for the radiation
$\g$, \be \delta  P_{{\rm intr}, \psi}=\delta  P_{{\rm intr},
\g}=0. \ee However, a scalar field  generically has a
non-vanishing non-adiabatic pressure perturbation \cite{KS}, \bea
\delta P_{{\rm intr}, \phi} &=& 2 V_{,\phi} \dot \phi
\left(\frac{\delta \rho_\phi}{\dot \rho_\phi} - \frac{\delta
\phi}{\dot \phi} \right) = -2 V_{, \phi} \left(\SM_\phi +
\frac{\dot \phi}{H} \zeta_\phi \right), \label{eq:nonadphi} \eea
where we have defined the gauge invariant variable \be \SM_\phi
\equiv \delta \phi +\frac{\dot \phi}{H} \Psi, \ee sometimes
referred to as the Sasaki and Mukhanov variable \cite{sasamukha}.
If the field $\phi$ is the only -- or the dominant -- component of
the universe, the intrinsic non-adiabatic pressure perturbation is
negligible on large scales (see e.g., \cite{Gordon}). However, in
our case the $\phi$ contributes only to a small amount of the
total energy and  $\delta P_{{\rm intr},\phi}$ cannot in general
be neglected.

The intrinsic non-adiabatic energy transfer function
is defined as \cite{wmu}
\be
\delta \QQ_{{\rm intr},\alpha} \equiv \delta \QQ_\alpha -
{\dot{\QQ}_\alpha \over \dot{\rho_\alpha}} \delta\rho_\alpha ,
\ee
where $\delta \QQ_\alpha$ is the perturbation of the energy transfer
function $\QQ_\alpha$ of Eqs.~(\ref{eq:defbackQa}-\ref{eq:defbackQc}).
It automatically vanishes if the energy
transfer function $\QQ_\alpha$  is only a function of the local
energy density $\rho_\alpha$ -- when neither the decay rate
nor the mass of the $\psi$-particles depend on the light field $\phi$.
For our three components $\psi$, $\gamma$, and $\phi$, we have
\begin{eqnarray}
\delta \QQ_{\rm{intr},\psi}&=& \rho_\psi \left\{
\left(C_M \frac{\ddot \phi}{\dot \phi}
+ \dot C_M - \frac{\dot \Gamma}{\dot \phi} \right) \left( \SM_\phi
+\frac{\dot \phi}{H} \zeta_\psi \right) +
C_M \left(\SM_\phi +\frac{\dot \phi}{H} \zeta_\phi  \right)
\right\} , \\
\delta \QQ_{\rm{intr},\g}&=&  -\rho_\psi \left\{ -\frac{\dot \Gamma}{\dot \phi}
\left( \SM_\phi +\frac{\dot \phi}{H} \zeta_\g \right)
+ \frac{\Gamma}{3H} \frac{  \dot \rho_\psi}{\rho_\psi }
{\cal S}_{\psi \g}
\right\}, \\
\delta \QQ_{\rm{intr},\phi}&=& - \rho_\psi \left\{
C_M \frac{\dot \phi}{3H}
\frac{\dot \rho_\psi}{\rho_\psi}  {\cal S}_{\phi \psi} +
\left(C_M \frac{V_{, \phi}}{\dot \phi}
- \dot C_M
\right)  \left(\SM_\phi +\frac{\dot \phi}{H} \zeta_\phi \right)
\right\}.
\end{eqnarray}
The relative non-adiabatic energy transfer function is
due to the presence of relative entropy perturbation
and is given by
\be
\label{deltaQrelalpha}
\delta \QQ_{{\rm rel},\alpha} \equiv
{\QQ_\alpha \dot\rho \over 2 H \rho} \left( \zeta - \zeta_\alpha \right)
= - {\QQ_\alpha \over 6 H \rho}  \sum_\beta \dot \rho_\beta
{\cal S}_{\alpha \beta}.
\ee
It vanishes if the background transfer function $\QQ_\alpha=0$.

Now we have all the ingredients to write the evolution equation of
the total curvature perturbation $\zeta$ on large scales, which is
obtained from Eq.~(\ref{eq:totalzetaevol}), \bea \dot \zeta &=&
\frac{H}{ \dot \rho^2} \left\{ \frac{1}{3} \dot \rho_\psi \dot
\rho_\g {\cal S}_{\psi \gamma} + c^2_\phi \dot \rho_\psi \dot
\rho_\phi {\cal S}_{\psi \phi} + \left( c^2_\phi - \frac{1}{3}
\right) \dot \rho_\g \dot \rho_\phi {\cal S}_{\gamma \phi}
\right\} \nonumber \\
&&+ \frac{2 H V_{, \phi} }{\rho +P} \left(\SM_\phi +\frac{\dot \phi}{H}
\zeta_\phi \right),
\label{eq:zetaevol}
\eea
and the three large scale evolution
equations of the uniform curvature perturbations $\zeta_\alpha$,
which are derived from Eq.~(\ref{eq:partialzetaevol}),
\bea
\dot \zeta_\psi &=&  H
\frac{\rho_\psi}{ \dot \rho_\psi} \left\{ \left(-C_M \frac{\dot \phi}{3H}
+ \frac{\Gamma}{3H} \right) \left(
\frac{\dot \rho_\gamma}{2 \rho} {\cal S}_{\gamma \psi}
+\frac{\dot \rho_\phi}{2 \rho}  {\cal S}_{\phi \psi} \right)
\right. \nonumber \\
&& \left. - \left(C_M \frac{\ddot \phi}{\dot \phi} +
\dot C_M  -C_\Gamma \Gamma \right)
\left( \SM_\phi + \frac{\dot \phi}{H} \zeta_\psi \right)
+ C_M \frac{\dot \rho_\phi}{\dot \phi^2}
\left( \SM_\phi +\frac{\dot \phi}{H} \zeta_\phi \right)
\right\}, \label{eq:master1}\\
\dot \zeta_\gamma &=&  H  \frac{\rho_\psi}{ \dot \rho_\gamma}
\left\{ \frac{\Gamma}{3H} \left[
\left( \frac{\dot \rho_\psi}{\rho_\psi} - \frac{\dot \rho_\psi}{2\rho}
\right) {\cal S}_{\psi \gamma} + \frac{\dot \rho_\phi}{2 \rho}
{\cal S}_{\g \phi} \right]
-C_\Gamma \Gamma
\left(\SM_\phi +\frac{\dot \phi}{H}\zeta_\g \right)  \right\}, \\
\dot \zeta_\phi &=&   H \frac{\rho_\psi}{ \dot \rho_\phi}
\left\{  -\frac{C_M \dot \phi}{3H}
\left[ \left(\frac{\dot \rho_\psi}{\rho_\psi} -
\frac{\dot \rho_\psi}{2 \rho} \right){\cal S}_{\psi \phi}
+ \frac{\dot \rho_\gamma}{2 \rho} {\cal S}_{\phi \g} \right]  \right.
\nonumber \\
&& \left.- \left( C_M \frac{V_{,\phi}}{\dot \phi} - \dot C_M
\right)
\left( \SM_\phi +\frac{\dot \phi}{H} \zeta_\phi \right)
\right\} -\frac{6 H^2 V_{,\phi}}{\dot \rho_\phi}
\left(\SM_\phi +\frac{\dot \phi}{H} \zeta_\phi \right),
 \label{eq:master3}
\eea where we have used $\dot \Gamma = C_\Gamma \Gamma \dot \phi$.
In order to close the system, we need the evolution equation for
the Sasaki-Mukhanov variable $\SM_\phi$ on large scales, namely,
\be \dot \SM_\phi =  \frac{\dot H }{H^2} \dot \phi \zeta
+\frac{\ddot \phi}{\dot \phi} \SM_\phi
 - \frac{\dot \rho_\phi}{\dot \phi^2}
\left(\SM_\phi +\frac{\dot \phi}{H} \zeta_\phi \right).
\label{eq:masterX} \ee
Equations~(\ref{eq:zetaevol}-\ref{eq:masterX}) are five coupled
first order differential equations -- one of them is redundant --
which can be solved in order to study the evolution of
perturbations during the inhomogeneous reheating. We did not make
any assumption for the intrinsic non-adiabatic pressure
perturbation of the field $\delta P_{{\rm intr}, \phi}$, which can
be found from $\SM_\phi$ and $\zeta_\phi$ [see
Eq.~(\ref{eq:nonadphi})], as well as for the coupling functions
$C_M$ and $C_\Gamma$. Thus these equations hold for any type of
functional dependency of the decay rate and mass on the light
field.

Some comments are in order here. The evolution equations of the
relative curvature perturbations $\zeta_\g$ and $\zeta_\phi$ are
sourced by terms proportional to $\rho_\psi$, although
$\zeta_\phi$ has an extra term proportional to its intrinsic
non-adiabatic pressure. After the decay of $\psi$, $\zeta_\g$ is
constant whereas $\zeta$ and $\zeta_\phi$ evolve only due to the
intrinsic entropy of the light field. If the coupling function
$C_M$ vanishes, Eqs.~(\ref{eq:zetaevol}-\ref{eq:masterX}) describe
the inhomogeneous inflaton decay, where $\psi$ represents the
oscillating inflaton and $\gamma$ the radiation produced by the
decay and reheating.

\section{Slow-roll limit}

As an application, we consider the limit where the scalar field $\phi$
is slow-rolling,
\be
\dot \phi/\sqrt{V} \simeq 0, 
\quad \ddot \phi/(H\dot \phi) \simeq 0, \label{eq:assumi}
\ee
and we assume that these conditions are maintained during all the
period of $\psi$ domination and decay.

In the slow-roll limit the scalar field varies very slowly in a
flat potential and its energy density is always negligible. In
this limit $c_\phi^2 \simeq -1$ and its intrinsic non-adiabatic
pressure perturbation vanishes from Eq.~(\ref{eq:nonadphi}), \be
\zeta_\phi \simeq -\frac{H}{\dot \phi} \SM_\phi. \ee Although the
variable $\zeta_\phi$ diverges in this limit, the Sasaki-Mukhanov
variable can still be used to discuss the scalar field
perturbation. Equations~(\ref{eq:zetaevol}-\ref{eq:masterX}) take
a very simple form, \bea \dot \zeta &=&  \frac{H}{\dot \rho}
\left\{ \dot \rho_\psi (\zeta_\psi -\zeta) +4 C_M H \rho_\psi
\SM_\phi
\right\}, \label{eq:mastersimple1}\\
\dot \zeta_\psi &=&  \Gamma
\frac{\rho_\psi}{ \dot \rho_\psi} \left\{
\frac{\dot \rho_\gamma}{2 \rho} (\zeta_\gamma -\zeta_\psi)
+C_\Gamma H  \SM_\phi \right\}, \\
\dot \zeta_\gamma &=&  H  \frac{\rho_\psi}{ \dot \rho_\gamma}
\left\{  \frac{\Gamma}{H}
\left( \frac{\dot \rho_\psi}{\rho_\psi} - \frac{\dot \rho_\psi}{2\rho}
\right) (\zeta_\psi - \zeta_\gamma) \right\}, \\
\dot \SM_\phi &=& 0.\label{eq:mastersimple2}
\eea

Also the background equations take a very simple form. In order to
solve them numerically, it is convenient to work in terms of
dimensionless quantities, the density parameters $\Omega_\alpha
\equiv \rho_\alpha/\rho$, with $
 \Omega_\psi +\Omega_\gamma =1$ ($\Omega_\phi \simeq 0$), and
the dimensionless reduced
decay rate \cite{wmu},
\begin{equation}
g\equiv\frac{\Gamma}{\Gamma+H} ,
\end{equation}
which varies monotonically from $0$ to $1$. In the slow-roll limit
of Eq.~(\ref{eq:assumi}) the background equations
(\ref{eq:dyn1}-\ref{eq:dyn4}) can then be written as an autonomous
system of first order differential equations,
\begin{eqnarray}
\label{eq:omegasprime}
 \Omega_\psi' &=&
 \left( \Omega_\g
-{g\over1-g}
\right) \Omega_\psi \label{din3} , \label{eq:backevol1}\\
 \Omega_\gamma' &=& \left( {g\over1-g} -\Omega_\g \right)
\Omega_\psi \label{din1},\\
 g' &=& {1\over2} g (1-g)
(4-\Omega_\psi), \label{eq:backevol4}
\end{eqnarray}
where the prime denotes differentiation with respect to the number
of $e$-foldings $N\equiv \ln a$. Since these equations are subject
to the constraint $\Omega_\psi+ \Omega_\g=1$ there are only two
independent dynamical equations whose solutions follow
trajectories in the  compact two-dimensional phase-plane
$(g,\Omega_\psi)$. One can find a detailed analysis of this
system, applied to the study of the curvaton model, in
Ref.~\cite{wmu}, where it is shown that close to the origin
$\Omega_\psi \simeq R_{\rm in} g^{1/2}$, with \be R_{\rm in}
\equiv \Omega_\psi \left. \left(\frac{H}{\Gamma}\right)^{1/2}
\right|_{\rm in}, \label{eq:Rin} \ee where the initial conditions
for this system are set at $t=t_{\rm in}$ such that $g \ll 1$. The
initial value $R_{\rm in}$ determines which trajectory is followed
in the two-dimensional phase-plane. For $R_{\rm in} \gtapprox 1$,
the massive species $\psi$ comes to dominate the universe before
decaying -- compare with Fig.~\ref{fig:r}. Indeed, the physical
interpretation of $R_{\rm in}$ is straightforward: If initially
$\Gamma \gg H \Omega_\psi^2$, the decay is almost instantaneous
and $\psi$ does not have the time to dominate.

The perturbation equations,
Eqs.~(\ref{eq:mastersimple1}-\ref{eq:mastersimple2}), can be
written in terms of the dimensionless background quantities
defined above and acquire a simple form \bea \zeta' &=&
\frac{\Omega_\psi}{4-\Omega_\psi} \left\{ \left(\frac{3-2g}{1-g}
\right) (\zeta_\psi -\zeta)
- 4 C_M \SM_\phi \right\}, \label{eq:manimani1}\\
\zeta_\psi ' &=& \frac{g }{3-2g} \left\{ \frac{1}{2} (4-\Omega_\psi)
(\zeta-\zeta_\psi) - C_{\Gamma} \SM_\phi \right\},
\label{eq:manimani2}\\
\SM_\phi'&=&0, \label{eq:manimani3} \eea where we have eliminated
the variable $\zeta_{\gamma}$, which is redundant. When $C_M=0$ we
recover the equations studied in \cite{Mazum} with $\psi$
representing the inflaton during its coherent oscillation. When
also $C_\Gamma=0$ we recover the equations studied in \cite{wmu}
with $\psi$ representing the curvaton.

\begin{figure}
\begin{center}
\includegraphics*[width=9cm]{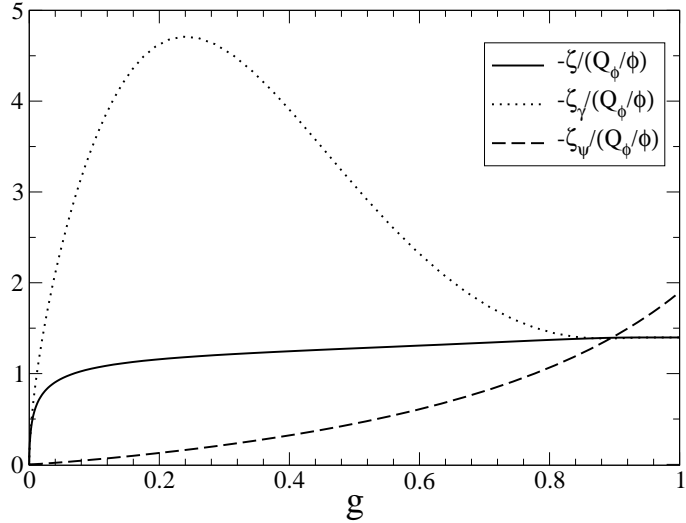}
\caption{Evolution of $\zeta$, $\zeta_\psi$, and $\zeta_\g$
normalized to the light field perturbation $- \SM_\phi/\phi$, as
a function of $g$,  for $R_{\rm in} =10$, where we have taken the
simplest case where the dependence of $\Gamma$ and $M$ is linear
in the field, $M \propto \Gamma \propto \phi$. Although it is not
visible on the ($\zeta_\bullet,g$)-plane, for $g \to 1$ the
uniform-density perturbations approach a fixed point attractor.}
\label{fig:example}
\end{center}
\end{figure}

\begin{figure}
\begin{center}
\includegraphics*[width=10cm]{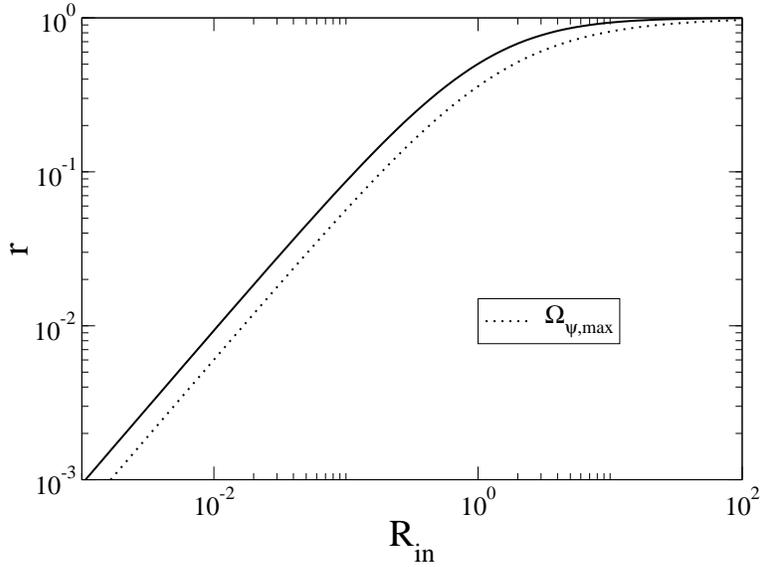}
\caption{The function $r$, parametrizing the conversion from an
isocurvature to a curvature perturbation, as a function of the
initial parameter $R_{\rm in}$ (solid line). This is compared to
$\Omega_{\psi,\rm max}$ (dashed line), the maximum value reached
by $\Omega_\psi$ before decaying as a function of $R_{\rm
in}=\Omega_\psi (H/\Gamma)^{1/2}|_{\rm in}$.} \label{fig:r}
\end{center}
\end{figure}

To calculate the final curvature perturbation on large scales, we
start with initial conditions at $g \ll 1$. By using the slow-roll
condition (\ref{eq:assumi}), from Eqs.~(\ref{eq:zetatot}) and
(\ref{eq:dyn1}-\ref{eq:dyn3}) the initial total curvature
perturbation is given by \bea \zeta_{\rm in}=\left.
\frac{1}{4-\Omega_\psi} (3 \Omega_\psi \zeta_\psi + 4 \Omega_\g
\zeta_\g - C_M \Omega_\psi \SM_\phi) \right|_{\rm in},
\label{eq:inizeta} \eea and in the two physical situations that we
discuss in this section it is negligible. Hence we shall consider
vanishing initial perturbations, $\zeta_{\rm in} \simeq
\zeta_{\g,\rm in} \simeq \zeta_{\psi, \rm in} \simeq 0$. Then we
numerically solve the system of
Eqs.~(\ref{eq:manimani1}-\ref{eq:manimani3}) and we evaluate the
perturbation variables for $g \to 1$. The late time solutions
approach a fixed point attractor, \bea \zeta =\zeta_\g &=&
-r(\alpha_\Gamma C_\Gamma + \alpha_M C_M )
\frac{\SM_\phi}{\phi } \equiv - r \alpha_\alpha \frac{\SM_\phi}{\phi } \label{eq:alpha} \\
&=& - r \left( \alpha_\Gamma \frac{\SM_\Gamma}{\Gamma}
+ \alpha_M \frac{\SM_M}{M} \right), \\
\zeta_\psi &=&  - r (\alpha_\Gamma +1/2) \frac{\SM_\Gamma}{\Gamma}
- r \alpha_M \frac{\SM_{M}}{M},\label{eq:effpsi} \eea where $r \le
1$ is a function of $R_{\rm in}$, whereas $\alpha_\Gamma$ and
$\alpha_M$ are constant which can be determined numerically. The
term $\alpha_\Gamma +1/2 $ in Eq.~(\ref{eq:effpsi}) comes from the
fixed point $\zeta_\psi' = 0$ for $\Omega_\psi=0, g = 1$ in
Eq.~(\ref{eq:manimani2}). An example of the evolution of
perturbations as a function of $g$ is given in
Fig.~\ref{fig:example}.

The behavior of $r$ as a function of $R_{\rm in}$ is illustrated
in Fig.~\ref{fig:r} and it is obtained by solving numerically
Eqs.~(\ref{eq:manimani1}-\ref{eq:manimani3}).
For large $R_{\rm in}$, $r=1$ and the efficiency of the mechanism
of conversion of the density perturbation of $\phi$ into  a
curvature perturbation is simply given by $\alpha_\phi$. For small
$R_{\rm in}$, $r \ll 1$ and the efficiency is considerably
reduced. In Fig.~\ref{fig:r} the maximum value reached by
$\Omega_\psi$ before the decay is also shown. For large $R_{\rm
in}$ one can approximate $r$ by this value. Note that $r$ as
defined here is the same as the one defined in \cite{curvaton4}
and computed in \cite{wmu} for the curvaton model. We numerically
checked that this is the case \footnote{The value of $r$ depends
only on the background solution and since for the curvaton and the
inhomogeneous reheating in the slow-roll limit this is the same,
also $r$ is the same.}.

We can represent the integration of
Eqs.~(\ref{eq:manimani1}-\ref{eq:manimani3}) as a transfer matrix
acting on the initial perturbations,
\be
 \label{eq:transfer}
\left(
\begin{array}{c}
\zeta \\ \SM_\phi /\phi
\end{array}
\right)_{\rm out}
 =
\left(
\begin{array}{cc}
1 & - r \alpha_\phi  \\
0 & 1
\end{array}
\right)
\left(
\begin{array}{c}
\zeta \\   \SM_\phi/\phi
\end{array}
\right)_{\rm in} . \ee Since the energy density of the light field
$\phi$ is always negligible, its perturbation represents an
entropy perturbation which is converted into a curvature
perturbation if $\alpha_\phi$ and $r$ are non-zero. Our task is
now to estimate the efficiency parameters $\alpha_\Gamma$ and
$\alpha_M$. Below we discuss two different physical situations.

\subsection{Inhomogeneous inflaton decay}

Here we discuss case (A) as mentioned in Sec.~II. The oscillating
inflaton is described by the fluid $\psi$ and $C_M=0$ is used.
Although fluctuating, the decay rate of the inflaton is time
independent in the limit (\ref{eq:assumi}). Initially, the
inflaton dominates the universe, $\Omega_\psi \to 1$ for $g \to
0$, whereas the radiation is negligible; thus we take $R_{\rm in}
\gg 1$ and $r=1$. The initial condition of $\zeta_\psi$ is given
by the initial perturbation of the inflaton field corresponding to
vacuum fluctuations. According to  \cite{gamma1} we assume that
the density perturbation of the inflaton field is negligible.
Therefore we have $|\SM_{\phi}/\phi_*| \gg |\zeta_{\psi,\rm in}|
=| \zeta_{\rm in}|$, where we have used Eq.~(\ref{eq:inizeta})
with $C_M=0$ and $\Omega_\g =0$ for the last equality, and
$\phi_*$ is the VEV of $\phi$ at horizon crossing. Solving
Eqs.~(\ref{eq:manimani1}-\ref{eq:manimani3}) with $R_{\rm in} \gg
1$ (inflaton domination) we find $\alpha_\Gamma =1/6$. This leads
to the result found in \cite{gamma1,Riotto,Mazum,Zalda}, \be \zeta
= -\frac{1}{6} \frac{\delta \Gamma}{\Gamma}, \ee valid on the
spatially flat slices $\Psi =0$.

\subsection{Inhomogeneous mass-domination and decay}

Here we discuss case (B) as mentioned in Sec.~II. Now $\psi$
is the massive particle species and both the mass $M$ and decay
rate $\Gamma$ depend on $\phi$. The radiation $\g$ is the product
of a previous reheating and it initially dominates the universe,
$\Omega_{\g, \rm in} \simeq 1$.

We start by discussing the condition for the massive particles to
dominate the universe before decaying. Assuming that the massive
species $\psi$ is sub-dominant when the initial conditions are
setup, the initial parameter $R_{\rm in}$ can be written as \be
R_{\rm in} \simeq \left( \frac{M^2}{M_{\rm Pl} \Gamma}
\right)^{1/2} g_r^{1/4} \frac{n_\psi}{s}, \label{eq:kkk} \ee where
$g_r$ is the number of relativistic degrees of freedom and
$n_\psi/s$ is the relic abundance of the $\psi$ particles when
they freeze-out ($s$ is the entropy density). In order to derive
Eq.~(\ref{eq:kkk}) we have used \cite{KolbTurner} \be \rho \sim
\rho_\g =\frac{\pi^2}{30} g_r T^4, \quad s =\frac{2 \pi^2}{45} g_r
T^3. \ee After freeze-out and when $\Gamma \ll H$, $n_\psi/s$ is
constant if $g_r$ is constant (which we shall consider
throughout). The massive species has time to dominate the universe
if $R_{\rm in} \ge 1$, which translates into
 \be \Gamma < \frac{M^2}{M_{\rm Pl}} g_r^{1/2} \left(
\frac{n_\psi}{s} \right)^2. \label{eq:dominopreciso} \ee

If  $n_\psi/s$ is order unity Eq.~(\ref{eq:conddomin}) is
recovered. However, it is interesting to try to plug some numbers
for the relic abundance in Eq.~(\ref{eq:dominopreciso}). We make
use of the analytic approximation for the relic abundance of
long-lived massive particles derived in \cite{JGK}. At high
temperature ($T \gg M$) $n_\psi \propto T^3$, whereas at low
temperature ($T \ll M$) the $\psi$ density is Boltzmann
suppressed, $n_\psi \propto T^{3/2} \exp(-M/T) $ so that if the
particles freeze-out when $T \gtapprox M$ then the $\psi$
abundance becomes very small. The initial equilibrium abundance is
maintained by annihilation of particles and antiparticles with
cross section $\sigma_A$ which we take to be independent of the
energy of the particles. In this case the abundance at freeze-out
is \cite{JGK} \be \frac{n_\psi}{s} \simeq \frac{100}{M M_{\rm Pl}
g_r^{1/2} \langle \sigma_A v \rangle}, \ee where $\langle \sigma_A
v \rangle$ is the thermal average of the total cross section times
the relative velocity $v$. On using this relation,
Eq.~(\ref{eq:kkk}) becomes independent of the mass $M$, \be R_{\rm
in} \simeq \frac{100}{g_r^{1/4}} \left( \frac{ M_{\rm Pl}}{\Gamma}
\right)^{1/2} \frac{ M_{\rm Pl}^{-2} }{\langle \sigma_A v \rangle
} . \ee This relation holds if the $\psi$-particles are
subdominant at the freeze-out. If a more detailed calculation is
performed one can see that the mass dependence enters only via a
logarithmic correction \cite{JGK}.
\begin{figure}
\begin{center}
\includegraphics*[width=9cm]{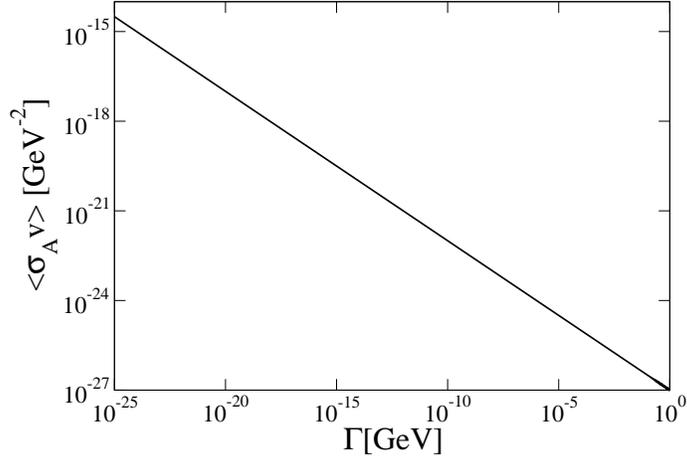}
\caption{Values of $\Gamma$ and $\langle \sigma_A v \rangle$ for
which $R_{{\rm in}}=1$. For values below this line the
$\psi$-particles come to dominate the universe. We have taken $g_r
\simeq 100$.} \label{fig:gamma}
\end{center}
\end{figure}
In Fig.~\ref{fig:gamma} we show the values of $\Gamma$ and
$\langle \sigma_A v \rangle$ for which $R_{\rm in} = 1$ and the
$\psi$-particles come to dominate. On requiring that the massive
particles decay before nucleosynthesis i.e.,  $\Gamma > (1 \ {\rm
MeV})^2/M_{\rm Pl}$, we find that only for $\langle \sigma_A v
\rangle < (100/g_r)^{1/2} (10^7 {\rm GeV})^{-2}$ does $\psi$
dominate the universe. In this case, the initial thermal
equilibrium by annihilation of particles and antiparticles must be
maintained by some gauge interaction much weaker than those of the
standard model. This excludes $\psi$ as being made by standard
model particles.

Let us now discuss the perturbations. The initial perturbation of
the radiation is left over by inflation and it is negligible. The
species  $\psi$ is initially in thermal equilibrium with $\g$,
$\zeta_{\psi, \rm in} = \zeta_{\g, \rm in}$, and
Eq.~(\ref{eq:inizeta}) implies $\SM_{\phi}/ \phi_* \gg
\zeta_{\g,\rm in} = \zeta_{\rm in}$. For the value of $\alpha_M$
we find numerically $\alpha_M = 4/3$. Thus, in general for a
$\phi$-dependent mass and decay rate \be \zeta = - r \left(
\frac{1}{6} \frac{\SM_\Gamma}{\Gamma} + \frac{4}{3}
\frac{\SM_M}{M} \right). \label{eq:general} \ee If the massive
particles dominate the universe before decaying, i.e., if $R_{\rm
in} \gg 1 $, $r=1$. For $M=$  constant we recover the result of
the inhomogeneous inflaton decay discussed in the preceding
section. If the mass depends on $\phi$ while $\Gamma$ does not, we
find \be \zeta = - \frac{4}{3} \frac{\SM_M}{M} . \label{eq:eccoci}
\ee Finally, if $M \propto \Gamma $ we find \be \zeta = -
\frac{3}{2} \frac{\SM_\Gamma}{\Gamma}, \ee a result nine times larger
than the one obtained in the case where only $\Gamma$ is
fluctuating.

In Ref.~\cite{gamma3} the perturbation generated by a varying mass
is derived with an analytic argument which yields $\zeta = (1/3)
(\SM_M/M) $, a result different from the one of
Eq.~(\ref{eq:eccoci}). This difference is due to the fact that in
our model, specified by the conservation equations
(\ref{eq:1}-\ref{eq:2}), the entropy perturbation stored in the
light field is important. Indeed, when $C_M \neq 0$, $\dot
\rho_\phi \neq 0$ and the light field perturbation contributes to
the relative non-adiabatic pressure $\delta P_{\rm rel}$ [see
Eq.~(\ref{eq:noadrel})] and sources the evolution of the total
curvature perturbation, as shown by Eq.~(\ref{eq:manimani1}).

\subsection{Comparison with the curvaton model}

It is worth discussing here the curvaton model, a mechanism of
generation of perturbations which has very similar properties as
the model discussed here. The aim is to stress their similarities
and compare their efficiency. The curvaton $\sigma$ is a scalar
field which is practically free during inflation and starts
oscillating after inflation (but before nucleosynthesis) during
the radiation era  when $\Gamma \ltapprox H \ltapprox m_\sigma$,
behaving as a non-relativistic fluid. If during the oscillations
it comes (close) to dominate the universe before decaying, its
perturbation $\zeta_\sigma$ is converted directly into curvature
perturbations, \be \zeta = r \zeta_\sigma.\label{eq:curvr} \ee We
can write the parameter $R_{\rm in}$ in terms of relevant
quantities. If we choose the initial time when the curvaton starts
to oscillate, at $H \sim m_\sigma$, on using $\rho_\sigma \sim
m_\sigma^2 \sigma^2$ we have, from the definition (\ref{eq:Rin}),
\be R_{\rm in} \simeq \frac{\sigma^2}{3 M_{\rm Pl}^2}
\left(\frac{m_\sigma}{\Gamma} \right)^{1/2}. \ee If the decay rate
is sufficiently smaller than the curvaton mass, the curvaton has
the time to dominate the universe before decaying. The conditions
for the curvaton domination during its oscillations are \be
 \left(\frac{\Gamma}{m_\sigma} \right)^{1/4}  M_{\rm Pl}
\ltapprox \sigma \ltapprox M_{\rm Pl}, \label{eq:curvacondimento}
\label{eq:constraintcurvaton} \ee where the last inequality
ensures that the curvaton does not dominate before starting to
oscillate.

Equations~(\ref{eq:manimani1}-\ref{eq:manimani3}) describe the
curvaton model if we set $C_M=C_\Gamma=0$ \cite{wmu} in which case
$\psi$ represents the curvaton during its oscillating phase, with
$\zeta_{\psi,\rm in} \equiv \zeta_{\sigma, \rm in} \gg \zeta_{\rm
in}$. By considering a massive curvaton we have \cite{curvaton2}
\be -H \frac{\delta \rho_\sigma}{\dot \rho_\sigma} = \frac{1}{3}
\frac{\delta \rho_\sigma}{\rho_\sigma} = \frac{2}{3} \frac{\delta
\sigma}{\sigma}. \ee For the last equality we have assumed that
the field remains overdamped until the Hubble parameter falls
below the curvaton mass, which is the case if
Eq.~(\ref{eq:curvacondimento}) is satisfied. Thus, from
Eq.~(\ref{eq:curvr}) we obtain \be \zeta = r \frac{2}{3}
\frac{\SM_\sigma}{\sigma} \equiv - r \alpha_{\sigma}
\frac{\SM_\sigma}{\sigma}. \label{eq:iden} \ee If we want to
compare the efficiency of the inhomogeneous reheating with that of
the curvaton model we must compare the efficiency parameter
$\alpha_{\sigma} = -\frac{2}{3} $ to $\alpha_\phi$ in
Eq.~(\ref{eq:alpha}), which varies according to the dependence of
the mass and decay rate on the light field $\phi$. However, in the
simplest case where $\Gamma \propto M \propto \phi$, $\alpha_\phi$
has opposite sign to $\alpha_\sigma$ and it is nearly twice
as larger.

We end this section with a comment. The physical situations
discussed in this section using
Eqs.~(\ref{eq:manimani1}-\ref{eq:manimani3}) assume that the mass
of $\phi$ must remain smaller than the Hubble parameter during the
whole process of inhomogeneous reheating. However, this relation
may be violated during the $\psi$ domination if $m_\phi \sim
\Gamma$ and $\phi$ starts to oscillate during this period. This
leads to a mix situation of curvaton/inhomogeneous reheating
scenario, where the perturbation of $\phi$ is converted into a
curvature perturbation via both the curvaton mechanism (the
curvaton being $\phi$) and the inhomogeneous reheating. Since the
sign of the efficiency parameters $\alpha_\bullet$ can be
different for the curvaton and inhomogeneous reheating models this
may lead to a compensation between them. An example of this
situation is given in \cite{Mazum}. The full calculation of the
resulting curvature perturbation in the inhomogeneous
mass-domination can be done starting from
Eqs.~(\ref{eq:zetaevol}-\ref{eq:masterX}).

\section{Observational constraint of the model}

Here we discuss the observational predictions of the inhomogeneous
reheating models: isocurvature perturbations and
non-Gaussianities.

\subsection{Isocurvature perturbations
in the mass-domination mechanism}

If the inflaton or the $\psi$-particles decay into species out of
equilibrium which remain decoupled from the radiation, we expect
isocurvature perturbations to be present into these species. These
can be correlated with the adiabatic perturbation. Here we
consider the case where the perturbations left over from inflation
are of the same order of magnitude as the perturbations produced
during the inhomogeneous mass-domination and decay. We define the
parameter $\kappa$ to quantify the relevance of the perturbations
left over from inflation, \be \kappa \equiv
r \alpha_{\phi} \frac{\SM_\phi/\phi}{\zeta_\chi}
= r \alpha_\phi \frac{V_{,\chi}}{3
H^2 \phi}, \label{eq:kappadef}\ee where $\chi$ is the inflaton,
$V_{,\chi}$ is the derivative of the inflaton potential,
and $H$ is the Hubble parameter, all
evaluated at horizon crossing during inflation. When $\kappa$ is
order unity or smaller, perturbations from inflation are
important. For chaotic inflation we have \be \kappa_{\rm chaotic}
\equiv 2r \alpha_\phi  \frac{M_{\rm Pl}^2}{\chi \phi}. \ee
Using an inflaton field which is $\sim 10 M_{\rm Pl}$ and an
efficiency $r \alpha_\phi \sim {\cal O}(1)$ we obtain $\kappa_{\rm
chaotic} \sim 0.2 M_{\rm Pl} /\phi$ so that perturbations from
inflation are important if the VEV of $\phi$ is sufficiently
large, $\phi \sim M_{\rm Pl}$.

The following analysis holds also for the curvaton model, although
the VEV of the curvaton during inflation should remain smaller
than the Planck mass [see Eq.~(\ref{eq:constraintcurvaton})] and
generally $\kappa \gg 1$. Thus, in the curvaton scenario it is not
likely that the inflaton and curvaton generated perturbations are
of the same order. The VEV of the light field $\phi$ does not need
to satisfy this constraint. Indeed, as discussed in \cite{gamma3},
if one wants to avoid that at high temperature the non-zero
density of the $\psi$-particles generates a large thermal mass for
$\phi$ -- which would make $\phi$ too heavy and would spoil the
simplicity of the mechanism -- we must require $\phi \sim M_{\rm
Pl}$. We are hence motivated to consider the possibility of
$\kappa$ being small, at least for chaotic inflation. In this case
perturbations from inflation may not be negligible with respect to
perturbations from the light field and a mix of the two may
survive. Isocurvature perturbations in the curvaton model are
discussed in \cite{curvaton4,GL} and in \cite{Moroi}, although
these groups considered a more general set of  possibilities than
what considered here and in \cite{GL,Moroi} they performed a
numerical analysis of the CMB data to constrain the curvaton
model. Here however we consider a different possibility i.e., that
the cold dark matter (CDM) is a relic left over both from the
inflaton and the $\psi$ decay.

We write the $\chi$ and $\phi$ quantum perturbations as $\hat
\SM_\chi = (H / \sqrt{2 k^3}) \hat a_\chi $ and $\hat \SM_\phi =
(H / \sqrt{2 k^3}) \hat a_\phi $, where the $\hat a_\alpha$'s
are independent normalized Gaussian random variables, obeying
$\langle \hat a_\alpha(\bk) \hat a_\beta(\bk') \rangle
=\delta_{\alpha \beta} \delta (\bk - \bk')$. After reheating, if
the relic product of the inflaton is decoupled from the product of
$\psi$, its uniform curvature perturbation is conserved and given
by \be \hat \zeta_{{\rm product~of~} \chi} =  \frac{3
H^2}{V_{,\chi }} \hat \SM_\chi = \hat a_\chi \zeta_\chi . \ee
From Eq.~(\ref{eq:transfer}) the uniform curvature perturbation of
the product of the $\psi$ decay can be written as \be \hat
\zeta_{{\rm product~of~}  \psi} = \frac{3 H^2}{V_{,\chi }} \hat
\SM_\chi - \frac{r \alpha_\phi}{\phi} \hat \SM_\phi = (\hat
a_\chi - \kappa \hat a_\phi  ) \zeta_\chi, \ee where the first
term in the right hand side comes from the initial perturbation
and the second from the fluctuations of $\phi$. Note that
$\kappa$, according to its definition (\ref{eq:kappadef}), is
scale dependent. We can write it as $\kappa(\bk) = \kappa_0
(k/k_0)^{\Delta n/2}$ where $\kappa_0$ is scale free and $k_0$ is
a reference pivot scale. The spectral index of $\kappa$, $\Delta
n$, can be expressed in terms of the difference between the
spectral indexes of $\zeta_\phi$ and $\zeta_\chi$, $\Delta n
\equiv n_\phi - n_\chi$, but here it is considered as a free
parameter.

To simplify the discussion we completely  neglect the baryons and
we concentrate on the CDM isocurvature mode, which is due to the
difference between the uniform curvature perturbations of the CDM
and radiation (e.g., photons and neutrinos). We start by assuming
that both the inflaton and the $\psi$-particles may decay in CDM
particles which are out of equilibrium at the temperature at which
they are produced. We define $f$ as the fraction of CDM, evaluated
just before nucleosynthesis, which is left over from the decay of
$\psi$. The rest of the CDM, $1-f$, is a relic of the inflaton
decay. Both the inflaton and the $\psi$-particles may decay into
radiation. The fraction of radiation that decays from $\psi$ is
proportional to the value of $\Omega_\phi$ at the decay, which we
assume to be $\Omega_{\phi, \rm max}$ defined in Sec.~IV.B. We
have seen there that this is very close to $r$. If $r \ll 1$,
$\psi$ remains negligible before decaying and cannot be
responsible for the radiation \footnote{Note that for mixed
perturbations the bounds on $r$ coming from the non-Gaussianities
discussed in the next section do not apply because adiabatic
perturbations are due to perturbations produced by inflation.
However, in the presence of correlated adiabatic and isocurvature
perturbations non-Gaussianity can be enhanced, see
\cite{RiottononG}}. However, $\psi$ can generate part of the CDM
leading to an uncorrelated isocurvature mode. This is constrained
by data: we have $f \kappa_0 < 0.28$ at $95 \%$ confidence level.
These bounds come from the numerical analysis of the WMAP data
made in \cite{Vali}, which assumes $-0.72 \le \Delta n \le 1.11$
and $k_0=0.05 \ {\rm Mpc}^{-1}$. The amount of CDM produced by the
decay of $\psi$ can be important only if the $\psi$ generated
perturbation is negligible. More interestingly, when $r\simeq 1$,
$\psi$ comes to dominate the universe and the totality of the
radiation comes from its decay product. In this case the adiabatic
and CDM isocurvature perturbations are correlated. The
intermediate case, that the radiation left over at nucleosynthesis
is created both by the inflaton and by the $\psi$ is not discussed
here. Indeed, the radiation thermalizes and it is difficult to
express its final perturbation in terms of primordial
perturbations.

We thus concentrate on the case $r \simeq 1$. Making use of the
notation of \cite{Vali} (see also \cite{AmendolaGordon}),
the adiabatic mode is described
by the comoving curvature perturbation in the radiation-dominated
era, ${\cal R}_{\rm rad} \equiv - \zeta_\gamma$, and the
isocurvature mode by  ${\cal S}_{\rm rad} \equiv 3(\zeta_c -
\zeta_\gamma)$, where $\zeta_\gamma$ and $\zeta_c$ are the uniform
curvature perturbations of the radiation (i.e., photons and
neutrinos) and of the CDM, respectively. Well in the radiation era
they are both constant and can be written as \bea
 {\cal R}_{\rm rad} &=& (\kappa \hat a_\phi - \hat a_\chi) \zeta_\chi, \\
 {\cal S}_{\rm rad} &=& 3  (1 - f) \kappa \hat a_\phi \zeta_\chi.
\eea According to \cite{Vali} we define the
dimensionless cross  correlation as
\be
\cos \Delta \equiv \left.
\frac{\langle  {\cal R}_{\rm rad} {\cal S}_{\rm rad}
\rangle}{(\langle{\cal R}_{\rm rad}^2  \rangle \langle{\cal
S}_{\rm rad}^2  \rangle  )^{1/2} } \right|_{k=k_0}
= \frac{  |\kappa_0 | }{\sqrt{ 1+
\kappa_0^2 }},
\ee
and the entropy-to-adiabatic ratio as
\be
f_{\rm iso}
\equiv \left. \left( \frac{ \langle {\cal S}_{\rm rad}^2 \rangle
}{\langle{\cal R}_{\rm rad}^2  \rangle  }
\right)^{1/2} \right|_{k=k_0}
= 3 (1 - f) \cos
\Delta. \ee These depend on two parameters, $f$ and $\kappa_0$. For
large values of $\kappa_0$ the  adiabatic perturbation is dominated
by the perturbation of the light field and the modes are totally
correlated ($\cos \Delta =1$) as found in \cite{GL}. More
generally  the  correlation is positive but can be small if the
inflaton perturbation becomes important.

\begin{figure}
\begin{center}
\includegraphics*[width=9cm]{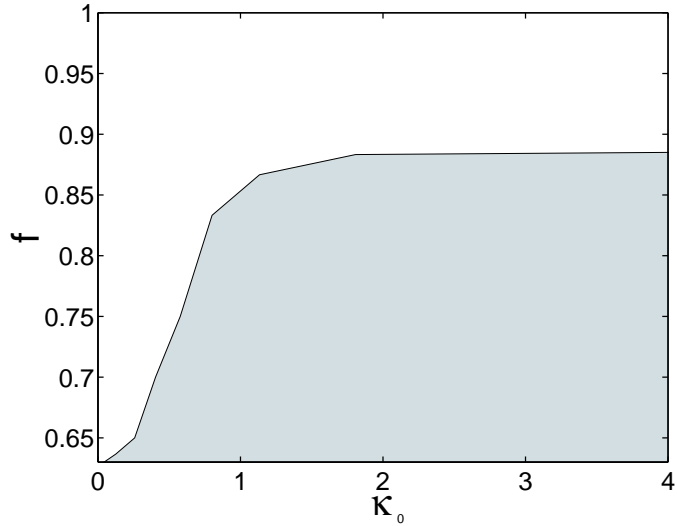}
\caption{Bounds on $f$, the fraction of CDM created by $\psi$,
as a function of $\kappa_0 \equiv
r \alpha_{\phi} \SM_\phi/(\phi \zeta_\chi) $ at $95 \%$
confidence level. The filled region is excluded.} \label{fig:iso}
\end{center}
\end{figure}

The entropy-to-adiabatic ratio is constrained by data and cannot
be too large. We can lower it by decreasing the amount of relic
CDM left over from inflation (i.e., by sending $f$ to 1) or by decreasing the amplitude of the
light field perturbation, i.e., the amplitude of $\kappa_{0}$. Since a
full analysis of the constraints imposed by the data on this model
is well beyond the scope of this paper, we just use the $95 \%$ confidence
level bounds on
the isocurvature mode coefficient $f_{\rm iso}$ as a function of $\cos \Delta$ for
correlated perturbations as given in \cite{Vali} (Fig.~1 in this reference) and we
show the bounds on $f$ as a function of $\kappa_0$ in
Fig.~\ref{fig:iso}.
Totally correlated perturbations are allowed only if the CDM is
almost completely created by $\psi$ ($f=1$). However, even for $\kappa \gg
1$, $\sim 10 \%$ of CDM created by the inflaton is allowed. The
signature of a correlated CDM isocurvature mode in the data can be
the signal that part of the CDM has been created before the decay
of $\psi$. These constraints apply equally well to the curvaton
model in the situation discussed here.

\subsection{Non-Gaussianities}

Until now we have assumed that the density perturbation of the
light field $\phi$ depends linearly on $ \SM_\phi/\phi$, which we
take to be a Gaussian variable. However, when the perturbation
$\SM_\phi$ is comparable to the average value $\phi$ -- which is
the case for small values of $\alpha_\phi$ --  the non-linear term
$(\SM_\phi/\phi)^2$ can be important and lead to a non-Gaussian
$\chi^2$ contribution in the spectrum of curvature perturbations
\cite{gamma3,Zalda}, \be \zeta = - r \alpha_\phi \left[
\frac{\SM_\phi}{\phi} - \frac{1}{2} \left( \frac{\SM_\phi}{\phi}
\right)^2 \right]. \label{eq:nongaus} \ee

The level of non-Gaussianity is conventionally specified by the
non-linear parameter  $f_{\rm NL}$ \cite{Kamiol,KomaSper}. We can
write the total uniform curvature perturbation as
\cite{Criminelli} \be \zeta =\zeta_g - \frac{3}{5} f_{\rm NL}
\zeta_g^2,\label{eq:nongaus2} \ee where $\zeta_g$ represents the
Gaussian contribution to $\zeta$. Using Eq.~(\ref{eq:nongaus}) the
prediction for the inhomogeneous reheating is \be f_{\rm NL} =-
\frac{5}{6 r \alpha_\phi}, \label{eq:ngalpha} \ee which is the
same as for the curvaton scenario once the replacement
$\alpha_{\phi} = - 2/3$ of Eq.~(\ref{eq:iden}) is used. However,
Eq.~(\ref{eq:ngalpha}) has to be taken with caution: In order to
precisely estimate the non linear parameter $f_{\rm NL}$ generated
by these models one has to study and solve the second order
perturbation equations as done in \cite{RiottoBartolo} and find
the second order correction to Eq.~(\ref{eq:general}). 

If we use Eq.~(\ref{eq:ngalpha}) we see that less efficiency in
the mechanism of conversion of perturbations means more
non-Gaussianities in the spectrum. If detected non-Gaussianity
could be the smoking-gun of models where perturbations are
produced with an `inefficient' mechanism of conversion. The WMAP
experiment has now put a limit on $f_{\rm NL}$ corresponding to
$-58 < f_{\rm NL} < 134$ at the $95 \%$ level \cite{WMAP4}, which
already excludes models with $-0.006 < \alpha_\phi < 0.014 $.
Planck will put a more sever constraint, $|f_{\rm NL}| \ltapprox
5$ \cite{KomaSper}, corresponding to $|\alpha_\phi| \gtapprox
1/6$.

Going back to the inhomogeneous reheating, in which the mechanism
of conversion of the density perturbation of $\phi$ into
a curvature perturbation is due to the fluctuations of  the decay
rate and $\Gamma \propto \phi$ we have $r \alpha_\phi \le 1/6$ and
the non-Gaussianity can be large, $-5 \le f_{\rm NL} < 0$. In
particular, if the non-relativistic species $\psi$ completely
dominates the universe before decaying ($r=1$) $f_{\rm NL}= - 5$
\cite{gamma3}, a value which is right in the ball-park of Planck
observations. However, the inhomogeneous mass-domination can be
much more efficient. If $M \propto \Gamma \propto \phi$, $r
\alpha_\phi \le 3/2 $ and thus $-5/9 \le f_{\rm NL} < 0$. If the
massive species dominates completely the universe before decaying
($r=1$) we have $ f_{\rm NL}=- \frac{5}{9} $, a much smaller value
than the one estimate in \cite{gamma3} and not observable by
future planned experiments.

\section{Conclusion}

In this paper we have studied the evolution of perturbations
during a phase dominated by massive particles whose mass and decay
rate can fluctuate in space and time. If the fluctuations are set
by the VEV of a light scalar field overdamped during inflation,
the isocurvature perturbation in the scalar field can be converted
into a curvature perturbation, and this can be the main mechanism
of generation of large scale perturbations for structure
formation. We have derived a set of perturbation equations that
can be used in full generality for any kind of dependence of the
mass and decay rate on the light field. Making use of these
perturbation equations we have recovered the results of
\cite{gamma1,Riotto,Mazum} for the inhomogeneous reheating with
varying decay rate. We have also discussed the condition for the
massive particles to dominate the universe before decaying. This
condition does not depend on their mass, but depends on the annihilation 
cross section and decay rate. Standard model massive particles cannot
dominate the universe. Furthermore, we have shown that when the
mass of the massive particles is allowed to vary, the mechanism of
conversion can be nine times more efficient. The final total
curvature perturbation is $\zeta = - (1/6) \delta \Gamma/\Gamma -
(4/3) \delta M/M $. This is our main result. Finally, we have
compared this with the curvaton model discussing differences and
similarities.

On the observational side we have discussed two possible
signatures of the mass-domination mechanism:
correlated adiabatic and isocurvature perturbations and
non-Gaussianities. If present, a cold dark matter isocurvature
perturbation provides some important information on the mechanism
of generation of the dark matter and on the vacuum expectation
values of the inflaton and light field during inflation. There are
non-Gaussianities generated by this mechanism, which are $\chi^2$.
In order to precisely compute them one has to study the evolution
of second order perturbations. In the limit where $f_{\rm NL}$ is large, 
by simply using
Eq.~(\ref{eq:ngalpha}) the non-linear parameter is $f_{\rm NL} =-
5/(6 r \alpha_\phi)$. When both the mass and the decay rate of the
massive particles fluctuate, due to the high efficiency the
non-Gaussianities can be much smaller than what is possibly
observable, $f_{\rm NL} \ll -5$.
\\

{\bf Acknowledgment}

It is a pleasure to thank Roberto Trotta for very fruitful
discussions and suggestions. I also acknowledge Ruth Durrer, David
Langlois, Karim Malik, and Jean-Philippe Uzan for very helpful
suggestions and comments, Robert Brandenberger for carefully
reading the manuscript, and Nicola Bartolo, Antonio Riotto and
Sabino Matarrese for drawing my attention on the importance of the
evolution of second order perturbations. Part of this work was
done at the Kavli Institute of Theoretic Physics. This research
was supported in part by the National Science Foundation under
Grant No. PHY99-07949 and by the Swiss National Science
Foundation.

\end{document}